%% file: QCLT.tex
\documentclass{amsart}
\newenvironment{nouppercase}{%
  \renewcommand{\uppercasenonmath}[1]{}}{}
\usepackage{enumerate,mathrsfs}
\usepackage{amsmath}
\usepackage{braket}
\usepackage{color}
\usepackage{graphicx}
\usepackage{tikz}
\usetikzlibrary{calc, shapes, positioning,backgrounds,arrows,automata,backgrounds,fit,decorations.pathreplacing}
\renewcommand{\subset}{\subseteq}

\DeclareMathOperator{\rank}{rank}
\DeclareMathOperator{\spn}{span}

\theoremstyle{definition}
\newtheorem{defn}{Definition}
\newtheorem{example}{Example}
\theoremstyle{theorem}
\newtheorem{theorem}{Theorem}

\DeclareMathOperator{\nn}{\mathbb{N}}

\title{\text{1D self-similar fractals with centro-symmetric Jacobians: asymptotics and modular data}}
\author{Radhakrishnan Balu$^{\dag}$}
\email{radhakrishnan.balu.civ@army.mil}

\begin{document}
\begin{nouppercase}
\maketitle
\end{nouppercase}
\begin{abstract}
We establish asymptotics of growing one dimensional self-similar fractal graphs, they are networks that allow multiple weighted edges between nodes, in terms of quantum central limit theorems for algebraic probability spaces in pure state. An additional structure is endowed with the repeating units of centro-symmetric Jacobians in the adjacency of a linear graph creating a self-similar fractal. The family of fractals induced by centro-symmetric Jacobians formulated as orthogonal polynomials that satisfy three term recurrence relations support such limits. The construction proceeds with the interacting fock spaces, T-algebras endowed with a quantum probability space, corresponding to the Jacobi coefficients of the recurrence relations and when some elements of the centro-symmetric matrix are constrained in a specific way we obtain, as the same Jacobian structure is repeated, the central limits. The generic formulation of Leonard pairs that form bases of conformal blocks and probablistic laplacians used in physics provide choice of centro-symmetric Jacobians widening the applicability of the result. We establish that the T-algebras of these 1D fractals, as they form a special class of distance-regular graphs, are thin and the induced association schemes are self-duals that lead to anyon systems with modular invariance.
\end{abstract}
\section {Introduction}
In a series of publications we characterized anyons, systems that are the building blocks of topological quantum computing, in terms of interacting fock spaces (IFS) and association schemes (AS) \cite { Radbalu2023, Radbalu2020, Radbalu2021}. The key idea is to identify modular invariance, that is central to rational conformal field theories \cite {Gann2005}, in a class of self-dual association schemes and cast them in terms of IFS that is a very generic way to treat quantum systems \cite {Accardi2017}. This AS-IFS description of anyons is concrete, as opposed to abstract modular tensor categories, in terms of related mathematical objects that encode fusion rules, conjugation, crossings in links, and braids. We can start with an IFS corresponding to graphs to set up fusion rules of anyons via the induced association scheme. The linear combination of the classes, that form the basis of the Bose-Mesner algebra, of the association scheme provides the matrix $W$ that encodes partition function of the spin system that can be set up on a graph. The $W$ matrix induces a commuting square from which we can derive a hyperfinite subfctor and the associated Termerley-Lieb algebra to describe the braidings of anyons. After we set up the background and establish the central limit theorem (CLT) we discuss the modular data associated with the fractal graphs.

Algebraic (quantum) probability spaces are a non-commutative generalization of classical probability spaces and in this paper we focus on graph induced $*-algebras$. We attach complex vector spaces to the vertices of the graph and the resulting algebra is endowed with a state which is a positive linear functional. The physical picture corresponds to a quantum particle whose configuration space is the vertices of the graph evolving under the influence of a magnetic field. As we are concerned with graphs an interacting Fock space (IFS) is a subconstituent algebra of adjacency matrices with a state, based on the fixed vertex, defined on it. In this work, our contribution is to construct interacting Fock spaces out of one dimensional self-similar graphs and establish central limit theorems. An IFS is a generalization of bosonic and fermionic Fock spaces, used in physics to describe the states of identical microscopic particles, that are $\mathbb{N}$-graded, each number correspond to number of particles, disjoint union of Hilbert spaces. In the context of association schemes, interacting Fock spaces arise as subconstituent algebras that have Hilbert space structures instead of vector spaces and further endowed with a state which is a real valued linear functional defined with respect to a fixed vertex of the algebra. The resulting structure is an algebraic probability space with adjacency matrices as random variables that are non commuting operators. This setting enables generalizing central limit theorems (CLT) of classical probability spaces to noncommutative settings with stochastic independence appropriately extended to the noncommutative context. We can start with an association scheme with a fixed number of classes and consider the operators of the corresponding T-algebra as quantum random variables. Now, we can grow this graph by increasing the diameter and thus consider a sequence of random variables that leads to the question of their limits along the lines of CLTs and denote them as QCLTs that have applications to quantum query complexity. In this work, we considered fractals whose adjacency matrices are irreducible tri-diagonal but the analysis can be extended to other fractals including 2D graphs. As quantum central theorems, they provide the limiting spectral distribution of the adjacency matrix of a growing graph, are relevant to routing in quantum networks and social networks represented as growing graphs, our results are significant in the context of fractal like graphs based information processing. Moreover, fractal graphs may be better suited for confining exotic phases of matter such as fractons \cite {Wen2018} our results have implications in topological quantum computations.

Let us define notions on algebraic probability space in the context of subconstituent algebras induced by distance regular graphs.
\begin {defn} Let $o$ be the fixed vertex of the subconstituent algebra $T$, of complex valued functions defined on the vertices, endowed with an inner product $\langle . , . \rangle$ and a pure state is a linear functional satisfying $$\rho_o (a) = \langle \delta_o , a\delta_o \rangle, a \in T$$.
The state and a notion of stochastic independence facilitate asymptotics of adjacency matrices of growing graphs via central limit theorems.
\end {defn}
\section {Association Schemes and T-algebras} 
An association scheme \cite {Radbalu2020} is a class of adjacency matrices of graphs with a set of $|\mathfrak{X}| = d$ vertices that encodes 1-distance, 2-distance, ..., d-distance adjacency of the graph.
Let $X$ be a (finite) vertex set, and $\mathfrak{X} = \{A_j\}_{j=0}^d$ be a collection of $X\times X$ $\{0,1\}$ matrices. The class $\mathfrak{X}$ is an \emph{association scheme} if the following hold:
\begin{enumerate}[(1)]
\item $A_0 = I$, the identity matrix;
\item $\sum_{j=0}^d A_j = J$, the all-ones matrix (In other words, the $1$'s in the $A_j$'s partition $X\times X$);
\item For each $j$, $A_j^T \in \mathfrak{X}$; and
\item For each $i,j$, $A_i A_j \in \spn\mathfrak{X}$.
\end{enumerate}
A \emph{commutative} association scheme also satisfies
\begin{enumerate}[(1)]
\setcounter{enumi}{4}
\item For each $i,j$, $A_i A_j = A_j A_i$.
\end{enumerate}
We note by $V = \mathbb{R}^X$ the vector space over $\mathbb{R}$ consisting of the column vectors with coordinates indexed by $X$ and all entries in $\mathbb{R}$.

Let us consider an association scheme $\{A_j\}_{j=0}^d$ that is commutative and from the spectral theorem we get an alternative basis $E_0,\dotsc,E_d$ of projections onto the maximal common eigenspaces of $A_0,\dotsc,A_d$. Since $\mathscr{A}$ is closed under the Hadamard product, element wise multiplication of matrices, forming a Bose-Mesner algebra, there are coefficients $q_{i,j}^k$ such that
\[ E_i \circ E_j = \frac{1}{|X|} \sum_{k=0}^d q_{i,j}^k E_k \qquad (0 \leq i,j \leq d). \]
The coefficients $q_{i,j}^k$ are called the \emph{Krein parameters} of the association scheme. This leads to a commutative hypergroup. Let $m_j = \rank E_j$, and define $e_j = m_j^{-1} E_j$. Then
\[ e_i \circ e_j = \frac{1}{|X|}\sum_{k=0}^d \left( \frac{m_k}{m_i m_j} q_{i,j}^k \right) e_k. \]

The dual notion to Krein parameters, the Intersection numbers $p^k_{ij}$ in terms of matrix product
$A_i \bullet A_j= \sum_{k} p^k_{ij}A_k$.
 For a distance-regular graph (ex: complete graphs, cycles, and odd graphs) intersection number describes the number of paths between a pair of k-distant vertices via i-distant plus j-distant paths is independent of the pair. For self-dual association schemes the Krein parameters and intersection numbers coincide.
 
 \begin {defn} Terwilliger algebras (T-algebras) \cite {Terwilliger1992} We can have the same Bose-Mesner algebraic structure using the idempotents $\{E_i\}$ and fixing a vertex $x$ of the graph in terms of $\{E_i(x)\}$. 
 
 This algebra is called Terwilliger algebra $T(x)$ with respect to $x$. We can define a T-module for this algebra, a subspace $W\subset V$ such that $BW\subset W, \forall B \in T$, that has a decomposition of orthogonal sum of irreducible modules.
\end {defn}
When the T-modules have dimension one with respect to any $x \in X$ then it is called thin and plays a role in charaterizing self-dual association schemes.

 There is a vast literature on T-algebras and we will connect them to IFS after introducing them with basic definitions \cite {Radbalu2021} as this correspondence helps to transfer techniques between the two independently developed fields.
\begin{example}
\item Let $G$ be a finite abelian group acting transitively on a finite set $X$. Then $G$ also acts on $X\times X$ through the action $g\cdot (x,y) = (g\cdot x, g\cdot y)$ for $g\in G$ and $x,y \in X$. Let $R_0,\dotsc,R_d \subset X\times X$ be the orbits for this action, numbered so that $R_0 = \{ (x,x) : x \in X\}$. (This is an orbit since $G$ acts transitively on $X$.) For each $j=0,\dotsc,d$, let $A_j$ be the $X \times X$ matrix with
\[ (A_j)_{x,y} = \begin{cases}
1, & \text{if } (x,y) \in R_j \\
0, & \text{otherwise.}
\end{cases} \]
Then, $\mathfrak{X} = \{A_j\}_{j=0}^d$ is an association scheme called translation scheme. It is commutative if and only if the action of $G$ on $X$ is \emph{multiplicity free}. In other words, the permutation representation of $G$ associated with its action on $X$ decomposes as a direct sum, of irreducibles, with no irreducible repeated up to unitary equivalence.
\end{example}

\section {Interacting Fock spaces}
Quantum probability based interacting Fock spaces generalize symmetric and anti-symmetric Fock spaces that have wide range of applications from quantum optics in physics to graph theory. These noncommutative spaces subsume classical probability spaces where there is one notion of stochastic independence by having several formulations that lead to various central limit theorems.  In quantum probability theory these independences are required to define graph products and based on the monadic operation different stochastic independence arise. In a quantum probability space $(\mathscr{A}, \phi)$ the usual commutative independence $(\phi(bab) = \phi(a)\phi(b^2); a,b \in \mathscr{A})$ such as the one assumed in quantum optics leads to conjugate Brownian motions (measured as quadratures) in the limit. The monotone independence $(\phi(bab) = \phi(a)\phi(b)^2; a,b \in \mathscr{A})$ that is relevant in quantum walks leads to arcsin-Brownian motion (double-horn distribution) aymptotically and the other two are free and Boolean independences not focused in this work. In the context of graphs the independence notions are defined in terms of products of graphs.

\begin {defn} \cite {Obata2007} An IFS associated with the Jacobi sequences $\{\omega_n\}, (\omega_m = 0) \Rightarrow \forall n \ge m, \omega_n = 0, \{\alpha_n\}, \alpha_n \in \mathbb{R}$ is a tuple $(\Gamma \subset \mathscr{H}, \{\Phi_n\}, B^+, B^-, B^\circ)$ where $\{\Phi_n\}$ are orthogonal polynomials and $B^\pm \Phi_n$ spans $\Gamma$, the subspace of the Hilbert space $\mathscr{H}$ which is a disjoint union of polynomials of degree $n$. The mutually adjoint operator $B^+, B^-$ and $B^\circ$ satisfy the relations
\begin {align*}
B^+ \Phi_n = \sqrt{\omega_{n+1}} \Phi_{n+1}. \\
B^- \Phi_n = \sqrt{\omega_n} \Phi_{n-1}; B^- \Phi_0 = 0. \\
B^\circ \Phi_n = \phi_n.
\end {align*}
\begin {equation} 
xP_n (x) = P_{n+1} (x) + \omega_n P_{n-1} (x) + \alpha_{n+1} P_n (x).
\end {equation}
\end {defn}
With the above IFS we can associate a graph with an adjacency matrix which is tridiagonal
$M = \begin{bmatrix} \alpha_1 & \sqrt{\omega_1}  \\
\sqrt{\omega_1} & \alpha_2 & \sqrt{\omega_2}  \\
& \sqrt{\omega_2} & \alpha_3 & \sqrt{\omega_3}  \\
 & & \ddots & \ddots & \ddots & \\
& & & \sqrt{\omega_{n-1}} & \alpha_n & \sqrt{\omega_n} \\
  & & & & \ddots & \ddots & \ddots 
\end {bmatrix}$
that has the quantum decomposition $T = B^+ + B^- + B^\circ$. The sequence $\{\Phi_n\}$ represents fixing a vertex and stratifying (partitioning based on distance from the fixed vertex) the graph with V set of vertices. For example, in the case of Spiderweb (Figure \ref {fig:spiderweb}) the origin is the fixed vertex at the center and stratification proceeds radially outwards.

Let us now consider the example of orthogonal polynomials of real numbers ($\mathbb{R}$) in detail and build an IFS.
\begin {example}
A probability measure $\mu$ on the real line $\mathbb{R}$ has a finite moment of order m if the relation holds: $\int_{-\infty}^{\infty} \overline{x}^m \mu (dx) < \infty$ and denoted by $M_m (\mu)$. On the other hand, a given sequence of real numbers form moments $\{ M_m \}$ of a probability measure if 
either all the elements are zero or only finitely many of them are non zero as $M_i > 0, 0 < i < m, M_j = 0, j > m$. This is a classical result in determinate moment problem. Let P and Q two complex valued polynomial functions in a single real variable and we can define an inner product between them as
\begin {align}
\mu (P) &= \int_{\mathbb{R}} P(x) \mu (dx). \\
\langle P, Q \rangle &= \mu(P^* Q).
\end {align}
This forms an algebra and we can obtain an orthogonal basis with respect to the measure $\mu$ and denote it as $\{ \Phi_n \}$. It can be shown that the orthogonal polynomials satisfy the following 3-diagonal relations and form an IFS \cite {Accardi2017}. 
\begin {equation}
x\Phi_n (x) = \Phi_{n+1} (x) + \alpha_n \Phi_n (x) + \omega_n \Phi_{n-1} (x).
\end {equation}
\end {example}
\begin {example}
For the bosonic (symmetric) Fock space we have $\omega_n = n; \alpha_n = 0$.
For the fermionic (anti-symmetric) Fock space the Jacobi parameters are $\omega_1 = 1; \omega_n =0, n > 1; \alpha_n = 0$.
\end {example}
\begin {example}
The q-deformed 1-mode IFS: For $q \ge -1$ the 1-mode IFS satisfying the conditions:
\begin {equation}
\omega_n = \begin {cases}
\sum_{k=0}^{n - 1} q^k, & \text { if q } > -1, \\
1, & \text { if q = -1 and n } \leq 1,  \\
0, & \text { if q = -1 and n } \geq 2.
\end {cases}
\end {equation}
is characterized the commutator $$aa^+ - qa^+ a = 1. $$
\end {example}

\newcommand{\D}{9} 
\newcommand{\U}{8} 

\newdimen\R 
\R=3.5cm 
\newdimen\L 
\L=4cm

\newcommand{\A}{360/\D} 

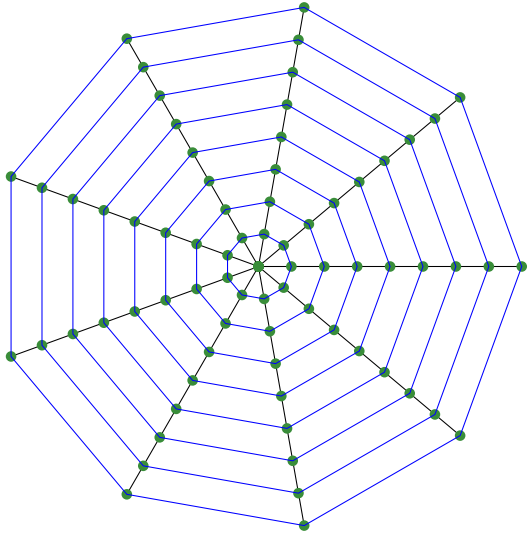
\begin{figure}[htbp]
 \centering
 \begin{tikzpicture}[scale=1]
  \path (0:0cm) coordinate (O); 

  \foreach \X in {1,...,\D}{
    \draw (\X*\A:0) -- (\X*\A:\R);
  }

  \foreach \Y in {0,...,\U}{
    \foreach \X in {1,...,\D}{
      \path (\X*\A:\Y*\R/\U) coordinate (D\X-\Y);
      \fill (D\X-\Y) circle (2pt) [fill={rgb:red,2;green,5;blue,2}];
    }
    \draw [blue, opacity=0.3] (0:\Y*\R/\U) \foreach \X in {1,...,\D}{
        -- (\X*\A:\Y*\R/\U) 
    } -- cycle;
  }
  \path (1*\A:\L) node (L1) {};
  \path (2*\A:\L) node (L2) {};
  \path (3*\A:\L) node (L3) {};
  \path (4*\A:\L) node (L4) {};
  \path (5*\A:\L) node (L5) {};
  \path (6*\A:\L) node (L6) {};
  \path (7*\A:\L) node (L7) {};
\end{tikzpicture}
\caption{Spiderweb Diagram - an example of a stratified graph on which an IFS can be defined.}
\label{fig:spiderweb}
\end{figure}

Let us state and outline the proof \cite {Obata2007} of QCLT for distance-regular graphs.

\begin {theorem} 
Let $\mathcal{G}^\nu = (V^\nu, E^\nu)$ be a growing distance-regular graph with an adjacency matrix $A_\nu$. Let us denote the degree as $\kappa(\nu)$ and assume the following conditions in terms of intersection numbers hold:
\begin {align*}
\omega_\nu &= \lim_{\nu \rightarrow \infty} \overline{\omega_\nu} &= \lim_{\nu \rightarrow \infty} \frac {p_{1, n -1}^n (\nu) p_{1, n}^n (\nu) } {\kappa(\nu)}. 
\\ 
\alpha_\nu &=  \lim_{\nu \rightarrow \infty} \overline {\alpha_\nu } &= \lim_{\nu \rightarrow \infty} \frac {p_{1, n -1}^{n - 1} (\nu)} {\sqrt(\kappa(\nu))}.
\end {align*}
Let $\Gamma_{\omega_n} = (\mathscr{G}, \{\Phi_n\}, B^+, B^-)$ be an interacting Fock space associated with $\{\omega_n\}$ and $B^o = \alpha_{N + 1}$ be the diagonal operator defined by $\{\alpha_n\}$, $N$ be the number operator. Then we have
\begin {equation}
\lim_{\nu \rightarrow \infty} = \frac {A^\epsilon _\nu} {\sqrt(\kappa(\nu))} = B^\epsilon, \epsilon = \{o, +, -\}.
\end {equation}
in the sense of stochastic convergence with respect to the pure state, i.e,
\begin {equation}
\lim_{\nu \rightarrow \infty} \langle \Phi^\nu _0, 
\frac {A_\nu^{\epsilon_m}}{\sqrt(\kappa(\nu))} \dots \frac {A_\nu^{\epsilon_1}}{\sqrt(\kappa(\nu))} \Phi^\nu _0\rangle = \langle \Psi_0, B^{\epsilon_m} \dots B^{\epsilon_1} \Psi_0 \rangle, \epsilon \in \{+, -, o\}, m = 1,2, \dots.
\end {equation}.
\end {theorem}
\begin {proof}
We have the following relations \cite {Obata2007}:
\begin {align*}
\frac {A^+ _\nu} {\sqrt(\kappa(\nu))} \Phi_n &= \sqrt{\overline{\omega_{n+1}}(\nu)}\Phi_{n+1}, n=0,1,2,\dots. \\
\frac {A^- _\nu} {\sqrt(\kappa(\nu))} \Phi_0 = 0; \frac {A^- _\nu} {\sqrt(\kappa(\nu))} \Phi_n &= \sqrt{\overline{\omega_{n}}(\nu)}\Phi_{n-1}, n=1,2,\dots. \\
\frac {A^o _\nu} {\sqrt(\kappa(\nu))} \Phi_n &= \overline{\alpha_n}(\nu) \Phi_n.
\end {align*}

From the above it follows that $\frac {A_\nu^{\epsilon_m}}{\sqrt(\kappa(\nu))} \dots \frac {A_\nu^{\epsilon_1}}{\sqrt(\kappa(\nu))} \Phi^\nu _0$ is a constant multiple of $\Phi^\nu _{\epsilon_1 + \dots \epsilon_m}$ and the constant is a finite product of $\omega_n (\nu)$ and $\alpha_n (\nu)$. Therefore, the left side of the limit exists.Moreover, since the actions of $A^\epsilon_\nu$ and $B^\epsilon_\nu$
on the number vectors are given by the Jacobi coefficients $\{\overline{\omega_n}\}$, $\{\overline{\alpha_n}\}$ and
$\{\omega_n\}$, $\{\alpha_n\}$, respectively, one may easily verify that the limit coincides with $\langle \Psi_0, B^\epsilon _m\dots B^\epsilon _1\Psi_0\rangle$.
\end {proof}

\begin {example} \cite {Obata2007} Let us consider a cyclic graph $C_{2N + 1}$ with $2N + 1$ vertices. Then, the intersection numbers required to obtain the limits of the theorem are:
\begin {equation*}
P_{1, n-1}^n (N) = \begin {cases} 1, & n = 1,2, \dots N,\\
 0, & \text {otherwise.}
\end {cases}
\end {equation*}

\begin {equation*}
P_{1, n}^{n - 1} (N) = \begin {cases} 2, & n = 1, \\
 1, & n = 2, \dots, N, \\
 0, & \text {otherwise.}
\end {cases}
\end {equation*}
\begin {equation*}
P_{1, n-1}^{n - 1} (N) = \begin {cases} 1, & n = N + 1,\\
 0, & \text {otherwise.}
\end {cases}
\end {equation*}
 It is easy to see that $\kappa = p^0_{11} = 0$ and so
 \begin {equation*}
\omega_n (N) = \begin {cases} 1, & n = 1, \\
1/2, & n = 2, \dots N, \\
 0, & \text {otherwise.}
\end {cases}
\end {equation*}
 \begin {equation*}
\alpha_n (N) = \begin {cases} 1/\sqrt{2}, & n = N + 1, \\
 0, & \text {otherwise.}
\end {cases}
\end {equation*}
\end {example}
This growing cyclic graph satisfies the conditions of the QCLT theorem. We will use similar techniques to constrain the Jacobians of the building blocks to get the asymptotics of self-similar graphs.

\begin {figure}
\usetikzlibrary {decorations.fractals} 
\begin{tikzpicture}[decoration=Cantor set,very thick]
  \draw [blue] decorate{ (0,0) -- (3,0) };
  \draw [blue] decorate{ decorate{ (0,-.5) -- (3,-.5) }};
  \draw [blue] decorate{ decorate{ decorate{ (0,-1) -- (3,-1) }}};
  \draw [blue] decorate{ decorate{ decorate{ decorate{ (0,-1.5) -- (3,-1.5) }}}};
\end{tikzpicture}
\caption{\label{fig:cantor} Cantor Set that is a self-similar graph}
\end {figure}
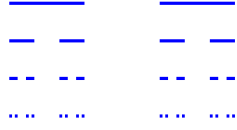

Recently Koheestani et al \cite {Koohestani2021} have established the QCLT for large family of distance-regular graphs with classical parameters in the Gibbs state. We extend the result further to self-similar weighted graphs (Figure \ref {fig:cantor}) in the pure state by constructing graphs that satisfy the conditions of the theorem above.

\subsection{Self-similar $p$ Laplacians on the half-integer lattice}\label{subsec2.1}

In an earlier work we considered a family of  self-similar Laplacians on the integers half-line and computed their spectra that are relevant to integer quantum Hall effects in physics \cite {Balu2022}. This class of Laplacians investigated in \cite{TeplyaevSpectralZeta2007} for the first time arise naturally when studying the unit-interval endowed with a particular fractal measure. Here, we focus on Laplacians without potentials  and consider the underlying graphs and their spectra.

In this context, we define the self-similar structure on the half-integer lattice with the origin serving the fixed vertex of our T-algebra. This self-similar structure describes a random walk on the half-line and gives rise to a class of self-similar probabilistic graph Laplacians $\Delta_p$ 

Let $\mathbb{Z}_+$ be the set of nonnegative integers and $\ell(\mathbb{Z}_+)$ be the linear space of complex-valued sequences $(f(x))_{x \in \mathbb{Z}_+}$. Let $p\in (0,1)$, for each $x\in \mathbb{Z}_+ \setminus \{0\}$, we define $m(x)$ to be the largest natural number $m$ such that $3^m$ divides $x$. For $f \in \ell(\mathbb{Z}_+)$ we define a \textit{self-similar Laplacian} $\Delta_p$ by,
\begin{align}\label{pLaplace}
	(\Delta_p f)(x) = \left\{\begin{array}{ll} f(0)-f(1), & \text{if}~x=0 \\
		f(x)-(1-p)f(x-1)-pf(x+1), &\text{if}~3^{-m(x)}x \equiv 1~\pmod 3 \\
		f(x) - pf(x-1)-(1-p)f(x+1), &\text{if}~3^{-m(x)}x \equiv 2~\pmod 3
	\end{array}\right..
\end{align}
We equip $\ell(\mathbb{Z}_+)$  with its canonical basis  $\{\delta_x\}_{x \in \mathbb{Z}_{+}}$ where 
\begin{equation}
\label{eq:canonicalBasis}
\delta_x(y)
 =
  \begin{cases}
  \  0     & \quad \text{if } x \neq y \\
  \  1 & \quad \text{if } x=y.
  \end{cases}
\end{equation}
 The matrix representation of $\Delta_p$  with respect to the canonical basis has the following Jacobi matrix
 
\begin{equation}
jacobi_{+,p} =
\begin{pmatrix}
1 & -1 & 0 & 0 & 0 & 0 & 0 & 0 & \dots  \\
p-1 & 1 & - p & 0 & 0 & 0 & 0 & 0 & \dots \\
0 & - p & 1 & p-1 & 0 & 0 & 0 & 0 & \dots \\
0 & 0 & p-1 & 1 & - p & 0 & 0 & 0 & \dots \\
0 & 0 & 0 & p-1 & 1 & - p & 0 & 0 & \dots \\
0 & 0 & 0 & 0 & - p & 1 & p-1 & 0 & \dots \\
0 & 0 & 0 & 0 & 0 & - p & 1 & p-1 & \dots \\
0 & 0 & 0 & 0 & 0 & 0 & p-1 & 1 &    \dots\\
\vdots & \vdots & \vdots & \vdots & \vdots & \vdots & \vdots & \vdots & \ddots
\end{pmatrix}.
\end{equation}
\text{ } \\
The case $p = \frac{1}{2}$ recovers the classical one-dimensional Laplacian (probabilistic graph Laplacian).  

\begin{defn}
 Let  $G_0 = (V_0, E_0)$ be the graph shown in Figure 
 We define the sequence of graphs $\{G_l \}_{l \in \nn}$ inductively. Suppose $G_{l-1}=(V_{l-1}, E_{l-1})$ is given for some integer $l \geq 1$, where $V_{l-1} =\mathbb{Z}_+ \cap [0,3^{l-1}]$. The graph $G_{l}=(V_{l}, E_{l})$ is constructed according to the following \textit{substitution rule}. We repeat the following steps for $i \in \{0,1,2\}$:
\begin{enumerate}
	\item  Insert a copy of $G_{l-1}$ between the two vertices $m_i$ and $m_{i+1}$ of the protograph shown in 
 in the following sense. We identify the vertex $0$ in $G_{l-1}$ with the vertex $m_i$ and similarly, we identify the vertex $3^{l-1}$ in $G_{l-1}$ with the vertex $m_{i+1}$.
	\item  We substitute the edges $(0,1)$ and $(3^{l-1},3^{l-1}-1)$ in  $G_{l-1}$ with the corresponding directed weighted edges as indicated in the protograph, see Figure 
\end{enumerate}
\end{defn}
\begin{figure}[htp]
\centering
\begin{minipage}{.5\textwidth}
\centering
\hspace*{-1cm} 
\resizebox{!}{!}{\input{G2_ToSubstt.tikz}}
\bigskip
\end{minipage}
\begin{minipage}[b]{.5\textwidth}
\centering
\hspace*{-4cm} 
\resizebox{!}{!}{\input{NeumannG3.tikz}}
\end{minipage}%
\caption{ Construction of self-similar graph from repeating units. (Top) A copy of the basic building block. The deleted edges correspond to the edges that are replaced when applying the substitution rule. (Bottom) The fractal graph is constructed by inserting the three copies of the building block in outer graph which is the 1D lattice 
While the vertices are labeled by the sequentially, the labeling of the edges represents the transition probabilities (off-diagonal entries in the self-similar Laplacian).}
\label{fig:G2Neumann}
\end{figure}
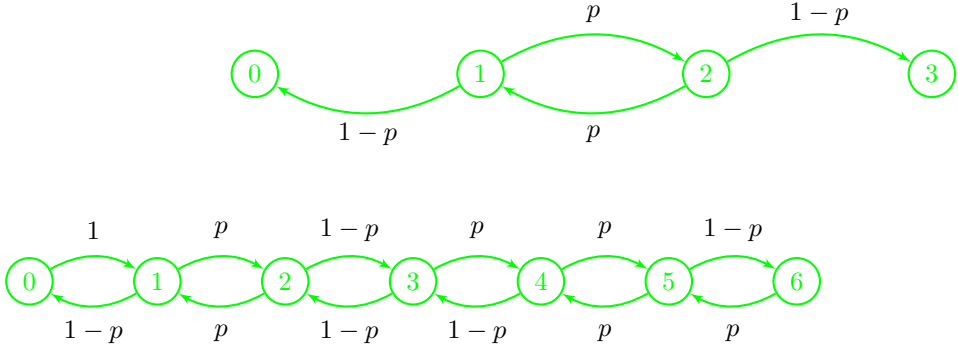
In a companion paper \cite {Balu2023} we discuss a family of centro-symmetric Jacobians and used them as building blocks for constructing self-similar graphs. There is a three-term recurrence relation for the building block and another one for the main graph. Let us consider the outer graph as that plays a role in the asymptotics.
\begin{align}
\label{eq:BasicJacobi}
jacobi_{cs} := \frac {1}{m}
\begin{pmatrix}
b(0) & a(1) & 0 &   \dots & 0   \\
a(n_0) & b(1) & a(2) &   \dots & 0  \\
0 & a(n_0-1) & b(2) &    \ddots &  \vdots  \\
\vdots & \vdots & \ddots & \ddots &   a(n_0)  \\
0 & 0 & 0 & a(1) & b(0)  
\end{pmatrix}
\end{align}
In the above we scale the centrosymmetric matrix by the inverse of the dimension of the matrix that will help with convergence later.
\begin{align}
\label{eq:recurrenceRelationsDiri}
\begin{cases}
    P^D_0(x)=1, \quad P^D_1(x) = x-b(1)\\
    P^D_k(x) = \big(x-b(k)\big) P^D_{k-1}(x) - a(k)a(n_0+1-k)P^D_{k-2}(x) , \quad k \in \{2,\dots, n_0-1\}.
  \end{cases}
\end{align}
Let us now state and establish the main result.
\begin {theorem} 
Let $\mathcal{G}^\nu = (V^\nu, E^\nu)$ be a growing 1D fractal (bidirectional network) with an adjacency matrix $A_\nu$. Let us denote the degree as $\kappa(\nu)$ and assume the following conditions hold:
\begin {align*}
\omega_\nu &= \lim_{\nu \rightarrow \infty} \frac {\overline{\omega_\nu}}{\kappa(\nu)}. 
\\ 
\alpha_\nu &=  \lim_{\nu \rightarrow \infty} \frac {\overline {\alpha_\nu }} {\kappa(\nu)}.
\end {align*}
Let $\Gamma_{\omega_n} = (\mathscr{G}, \{\Phi_n\}, B^+, B^-)$ be an interacting fock space associated with $\{\omega_n\}$ and $B^o = \alpha_{N + 1}$ be the diagonal operator defined by $\{\alpha_n\}$, $N$ be the number operator. Then we have
\begin {equation}
\lim_{\nu \rightarrow \infty} = \frac {A^\epsilon _\nu} {\sqrt(\kappa(\nu))} = B^\epsilon, \epsilon = \{o, +, -\}.
\end {equation}
in the sense of stochastic convergence with respect to the pure state, i.e,
\begin {equation} \label{eq:limitEquation}
\lim_{\nu \rightarrow \infty} \langle \Phi^\nu _0, 
\frac {A_\nu^{\epsilon_m}}{\sqrt(\kappa(\nu))} \dots \frac {A_\nu^{\epsilon_1}}{\sqrt(\kappa(\nu))} \Phi^\nu _0\rangle = \langle \Psi_0, B^{\epsilon_m} \dots B^{\epsilon_1} \Psi_0 \rangle, \epsilon \in \{+, -, o\}, m = 1,2, \dots.
\end {equation}.
\end {theorem}
\begin {proof}
We can rewrite the above equation \eqref {eq:recurrenceRelationsDiri} as to get the Jacobi coefficients:
\begin {align*}
x P^D_{k-1}(x) &=  b(k) P^D_{k-1}(x) + P^D_k(x) + a(k)a(n_0+1-k)P^D_{k-2}(x). \\
\omega_k &= a(k)a(n_0 + 1 -k). \\
\alpha_{k + 1} &= b(k).
\end {align*}

The $a(i), b(j)$ terms in repeating units are bounded by $1$ and there are only finitely many terms (finite moments) in equation \eqref {eq:limitEquation} so the limit exists for the outer graph. Since, the elements $a(i)$ are probabilities the scaling constant $m$ is not required. If we build the graph with probabilistic Laplacian then it is clear that we have the limits for the Jacobi coefficients as the numerators are probabilities and less than one and the denominator is $\kappa = 4$,  and the QCLT theorem holds. The limiting measure can be obtained by the spectral decimation method \cite {Balu2022} as the methods of applying continued fractions \cite {Obata2007} are difficult in general for an arbitrary distance-regular graph.

Another class of systems can be constructed starting from any Leonard pairs \cite {Paul2003} and taking the centro-symmteric Jacobians out of the pairs. For example the pairs where $d$ is a any non-negative integer:
\begin {equation}
A = \frac {1}{m}\begin{pmatrix} 0 & d & & & & 0 \\
1 & 0 & d - 1 & & & \\
 & 2 & . & . \\
 & & . & . & . \\
 & & & . & . & 1 \\
 0 & & & &  d & 0
\end {pmatrix}; B = diag (d, d - 2, d - 4, \dots, -d);
\end {equation}
When we normalize the above centro-symmetric matrix and build the self-similar graph then again we will have QCLT with spin Leonard pairs form the bases related by Krawtchouk polynmoials \cite {Kaul1994}.

In the above examples we can replace the diagonal elements of the centro-symmetric Jacobian all zeros with an integer less than the degree $\kappa = 4$ and we will still have convergence.
\end {proof}
It is interesting to note that the adjacency matrix of our self-similar graphs are irreducible tridiagonal ( each entry on the subdiagonal is nonzero) with nonnegative entries and thus has a bidirectional path and is described by a Q-polynomial \cite {Paul2011}. 

\section{Modular Data of Rational Conformal Field Theory}
There is an abstract way to define association schemes as character algebra that are in one-to-one correspondence with fusion algebras of conformal field theories such as the $SU(2)_k$ based ones\cite {Bannai1993}.  Gannon set up an axiomatic framework to elaborate this correspondence and identified several fusion algebras including association schemes in the language of conformal field theory \cite {Gann2005}. The central idea is that the modular data or invariance present in these algebras that is described in terms of $S-$matrix and a diagonal matrix $T$ as $(ST)^2 = S^2$. In the case of association schemes, modular invariance is satisfied by self-duals that induce a matrix $W$ which is a linear combination of the classes of the scheme. The matrix $W$ called a spin model encodes crossings in a knot by respecting all the three Reidemeister moves, and represents the statistical partition function and further induces a subfactor. Thus, it describes conjugation, fusion, link invariant, braiding of an anyon in a single mathematical object. In a recent publication we have detailed the modular data for the self-dual Hamming association scheme \cite {Radbalu2023} and established the subfactor induced TQFT and the Leonard pair based conformal blocks correspondence. There is another way to characterize self-dual  association schemes corresponding to distance-regular graphs in terms of Leonard pairs and T-modules that are thin which is a necessary and sufficient condition \cite {Curtin1999, Terwilliger2022}. As our construction of the families of 1D fractals are based on spin Leonard pairs of type III (Krawtchouk) as the centro-symmetric Jacobian for the repeating units the resulting distance-regular graphs are self-duals and thus support modular data.
 
\section {Summary and Conclusions}
We investigated graphs with weighted edges and endowed with self-similar fractal structures. We derived the QCLT for a family of graphs in pure state by constraining the centro-symmetric Jacobian that generate the fractals. The class of fractal graphs considered here lead to self-dual association schemes and thus encode modular invariance of RCFTs. This analysis sets the stage for exploring QCLT for fractals in coherent states that are relevant in physics and more general fractals in 2D such as the Sierpinski gasket.
\
\section {Acknowledgement} The author is grateful to Paul Terrwilliger for suggesting the family of Leonard pairs that have centro-symmetric jacobians.

\section {Declarations}
\textbf {Funding and/or Conflicts of interests/Competing interests:} The  funding information is not applicable and there are no conflicting or competing interests associated with this manuscript.\\
\textbf {Data availability:} This manuscript has no associated data.

\bibliographystyle{abbrv}

\end{document}

%% file: G2_ToSubstt.tikz.tex
\begin{tikzpicture}

\tikzset{vertex/.style = {shape=circle,green,thick,draw,minimum size=1.5em}}
\tikzset{edge/.style = {->,> = latex', green,thick}}
\node[vertex] (0) at  (0,0) {0};
\node[vertex] (1) at (3.,0) {1};
\node[vertex] (2) at  (6,0) {2};
\node[vertex] (3) at (9,0) {3};

\node at (1.5,-0.8){$1-p$};

\node at (4.5,0.8){$p$};
\node at (4.5,-0.8){$p$};

\node at (7.5,0.8){$1-p$};

\draw [red] [edge] (1) to[bend left] (0);

\draw[edge] (1)  to[bend left] (2);
\draw[edge] (2) to[bend left] (1);

\draw[edge] (2)  to[bend left] (3);

\end{tikzpicture}

%% file: NeumannG3.tikz.tex
\begin{tikzpicture}[scale=0.85]

\tikzset{vertex/.style = {shape=circle,green,thick,draw,minimum size=1.5em}}
\tikzset{edge/.style = {->,> = latex',green,thick}}
\node[vertex] (0) at  (0,0) {0};
\node[vertex] (1) at (2,0) {1};
\node[vertex] (2) at  (4,0) {2};
\node[vertex] (3) at (6,0) {3};
\node[vertex] (4) at  (8,0) {4};
\node[vertex] (5) at (10,0) {5};
\node[vertex] (6) at  (12,0) {6};


\node at (1.,0.8){$1$};
\node at (1.,-0.8){$1-p$};

\node at (3,0.8){$p$};
\node at (3,-0.8){$p$};

\node at (5,0.8){$1-p$};
\node at (5,-0.8){$1-p$};

\node at (7,0.8){$p$};
\node at (7,-0.8){$1-p$};

\node at (9,0.8){$p$};
\node at (9,-0.8){$p$};

\node at (11,0.8){$1-p$};
\node at (11,-0.8){$p$};


\draw[edge] (0)  to[bend left] (1);
\draw[edge] (1) to[bend left] (0);

\draw[edge] (1)  to[bend left] (2);
\draw[edge] (2) to[bend left] (1);

\draw[edge] (2)  to[bend left] (3);
\draw[edge] (3) to[bend left] (2); 

\draw[edge] (3)  to[bend left] (4);
\draw[edge] (4) to[bend left] (3); 

\draw[edge] (4)  to[bend left] (5);
\draw[edge] (5) to[bend left] (4); 

\draw[edge] (5)  to[bend left] (6);
\draw[edge] (6) to[bend left] (5);

\end{tikzpicture}